\documentclass[aps,pre,reprint,10pt,superscriptaddress]{revtex4-2}  
\usepackage{graphicx}
\usepackage{amsmath,amssymb,bm, xcolor}
\usepackage[top=1.5cm, bottom=1.5cm, left=1.7cm, right=1.7cm]{geometry}  
\usepackage{svg} 
\usepackage{xcolor}
\usepackage{float}
\usepackage{placeins}
\svgpath{{figures/}} 
\svgsetup{inkscapelatex=false} 
\usepackage{subcaption}
\definecolor{bottlegreen}{RGB}{0,106,78}
\usepackage{changes}
\usepackage[T1]{fontenc} 
\usepackage{braket}
\usepackage{siunitx}
\usepackage{comment}
\usepackage{titlesec}
\titlespacing*{\section}{0pt}{2ex minus .2ex}{1ex minus .2ex}
\titlespacing*{\subsection}{0pt}{1.5ex minus .2ex}{1ex minus .2ex}
\titlespacing*{\subsubsection}{0pt}{1ex minus .2ex}{0.5ex minus .2ex}

\usepackage[colorlinks=true, citecolor=blue]{hyperref}  
\begin{document}

\title{Quantum control of spin qubits using SOT-driven nanomagnets}

\author{Aniruddha Chakraborty}
\altaffiliation{These authors contributed equally to this work.}
\affiliation{Department of Mechanical and Nuclear Engineering, College of Engineering, Virginia Commonwealth University}

\author{Kanishk Modi}
\altaffiliation{These authors contributed equally to this work.}
\affiliation{Department of Physics, Indian Institute of Technology Bombay, Powai, Mumbai - 400076, India}

\author{Dhritiman Bhattacharya}
\affiliation{Department of Electrical \& Computer Engineering, Henry M. Rowan College of Engineering, Rowan University}

\author{Uditendu Mukhopadhyay}
\affiliation{Department of Physics, Indian Institute of Technology Bombay, Powai, Mumbai - 400076, India}
\affiliation{Centre of Excellence in Quantum Information, Computing, Science \& Technology, Indian Institute of Technology Bombay, Powai, Mumbai - 400076, India}

\author{Suddhasatta Mahapatra}
\email[Suddhasatta Mahapatra: ]{suddhasatta.mahapatra@gmail.com}
\affiliation{Department of Physics, Indian Institute of Technology Bombay, Powai, Mumbai - 400076, India}
\affiliation{Centre of Excellence in Quantum Information, Computing, Science \& Technology, Indian Institute of Technology Bombay, Powai, Mumbai - 400076, India}

\author{Jayasimha Atulasimha}
\email[Jayasimha Atulasimha: ]{jatulasimha@vcu.edu}
\affiliation{Department of Mechanical and Nuclear Engineering, College of Engineering, Virginia Commonwealth University}

\date{\today}

\begin{abstract}
Spin rotation (SR) is an essential capability for realization of single and two-qubit gates in spin quantum computing (SQC) architectures. To perform SR, resonant AC magnetic fields are either generated by microwave current pulses fed to an antenna, or voltage pulses applied to a gate, in presence of an inhomogeneous Zeeman field. While the former approach is limited by gate-speed and site-selectivity of SR, the latter adds to the decoherence of the spin qubits. Here, we propose an alternative technique for driving high-speed SR without compromising the qubit coherence, by employing spin-orbit-torque (SOT)-driven nanomagnets to produce oscillating magnetic fields, locally at the qubit site. The proposed scheme is highly energy-efficient, scalable, and compatible with the CMOS fabrication technology.
\end{abstract}


\maketitle

\section{Introduction}

 In recent implementations of spin quantum computing architectures, with spin qubits hosted by semiconductor quantum dots, electric dipole spin resonance (EDSR) has emerged as a reliable technique for realization of fast and fiducial single-qubit gates \cite{MM_academia1, MM_academia2, scaling_eg1, scaling_eg2, scaling_eg3}. EDSR relies upon local application of microwave (MW) pulses at the qubit site, in the presence of  magnetic field gradients, the latter typically provided by micromagnets (MMs) \cite{MM_Tokura, MM_Obata}. With EDSR, Rabi frequencies ($f_R$) exceeding 70 MHz \cite{maurand_CMOS}, and gate fidelities of up to 99.9\% \cite{yoneda2018quantum}, have been demonstrated in silicon spin qubits. The technique further allows selective addressing of spin qubits in an array, since the spatially inhomogeneous  magnetic field due to the MMs yields sufficiently distinct Larmor frequencies ($f_{L}$) of neighboring qubits. However, the magnetic field gradient of the MMs is also responsible for significantly high decoherence rates of spin qubits, due to an artificially induced spin-orbit coupling.

 On the other hand, initial schemes of SQC relied upon the more traditional electron spin resonance (ESR) technique, wherein the oscillating magnetic field, resonant with $f_L$, is provided by MW-pulses fed to an on-chip antenna \cite{ESR_Koppens, ESR_Veldhorst2014}. In this approach, $f_R$ was restricted to few MHz, since high MW-currents led to heating of the chip \cite{peetroons2025esr, ESR_Morello}. Additionally, with a relatively large form factor of the on-chip antenna, and a homogeneous externally-applied B-field, the selectivity of spin rotation by ESR also proved to be challenging.

In this work, we propose an alternative scheme of realizing fast, high-fidelity and frequency-selective spin rotation, which allows the qubit decoherence to be significantly suppressed, without compromising the gate speed. The approach makes use of SOT-driven nanomagnets (NM) to directly generate oscillating B-fields, akin to the ESR technique. Yet, the stray B-field due to the nano-magnets enables selective addressing of individual qubits, while the low gradients ensure suppressed qubit decoherence. Additionally, the scheme is energy-efficient, and fully-compatible with the standard CMOS technology.

\begin{figure}[ht]
    \centering
    
    \begin{subfigure}{1.0\linewidth}
        \centering
        \includegraphics[width=1\linewidth]{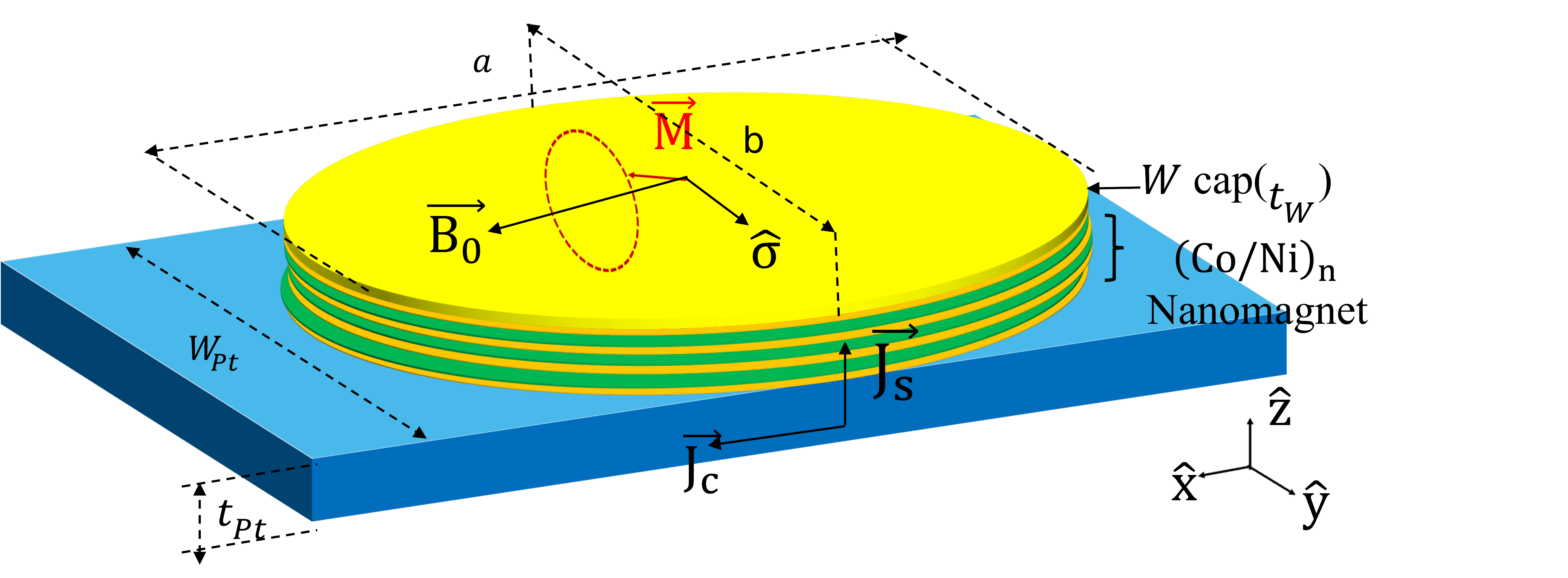}
        \caption{}
        \label{fig:SOT}
    \end{subfigure}

    \begin{subfigure}{1.0\linewidth}
    \centering
        \includegraphics[width=1\linewidth]{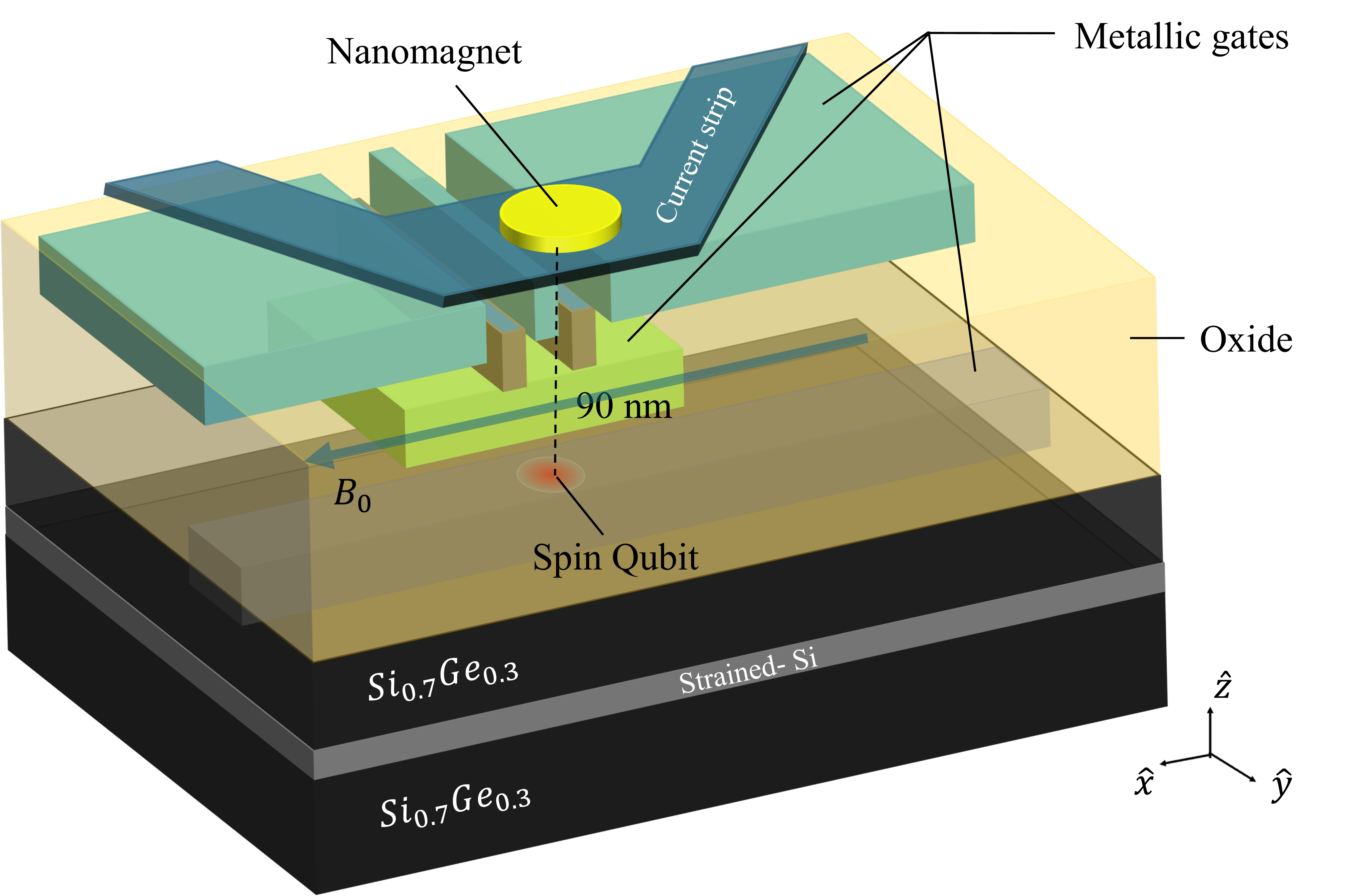}
        \caption{}
        \label{fig:SQ}
    \end{subfigure}

    \caption{\textbf{(a)} Schematic representation of magnetization dynamics in the NM. $B_0$ is the externally applied magnetic field, and the magnetization of the NM ($\Vec{M}$) is shown precessing around $\Vec{B_0}$. \textbf{(b)} Schematic representation of the spin qubit device, showing the gate layers, the SOT strip-line, the NM, and the QD hosting the spin qubit in the strained-Si QW of the Si/SiGe heterostructure. The NM is positioned directly above the QD, 90 nm above the Si-QW plane.}
    \label{SQ_SOT}
    
\end{figure}

The concept of generating a MW B-field, with a SOT-driven NM, is elucidated in Fig. \ref{fig:SOT}. An in-plane charge current through the thin heavy-metal (HM) layer yields an out-of-plane spin current, due to the spin-Hall effect. This pure spin-current exerts a damping-like torque on the magnetization of the ferromagnetic (FM) layer on top, thereby driving it out of equilibrium. An AC charge current through the HM layer leads to magnetization precession in the FM layer, thereby generating an oscillating stray B-field, which can enable the spin-rotation of a nearby qubit. This approach of generating the oscillating B-field is very promising in the context of selective addressability of spin qubits, since the dipolar field of the FM is highly localized (decaying as $r^{-3}$, where $r$ denotes the distance away from the NM). Although nano-scale magnetic devices have been studied for spin control \cite{NMniknam2022,Skyrmion_Fahim2024} and nanoscale and mesoscale magnetic devices have been used for manipulation of NV centers in diamond \cite{NVDW_Nathan2023,NVMTJ_Gerald2024,NVSAW_Chowdhury2024, NVSpinwave_Jung2024}, this work suggests the applicability of the concept for electrically-driven spin qubits in semiconductors. 


For our simulations, we integrate the nanomagnet device of Fig. \ref{fig:SOT} with a typical $\mathrm{Si/Si_{1-x}Ge_{x}}$ heterostructure used for spin qubit architectures. A 10-nm strained-Si quantum well (QW) is sandwiched between a 30-nm $\mathrm{Si/Si_{0.7}Ge_{0.3}}$ spacer on top and a 300-nm $\mathrm{Si/Si_{0.7}Ge_{0.3}}$ barrier at the bottom, with a 2-nm Si cap at the surface (Fig. \ref{fig:SQ}). 
The quantum dots (QD) hosting the individual spins are electrostatically induced, by applying appropriate voltages to the tri-layer metallic gates at the surface of the heterostructure, and embedded within a sufficiently-thick oxide layer. The oxide layer provides electrical isolation between the different gate levels, and with the NM as well, while allowing careful positioning of the latter with respect to the qubit plane (Fig. \ref{fig:SQ}). In our simulations, the qubit plane is defined at the Si QW. We assume the g-factor to be 2, for our choice of the $\mathrm{Si/Si_{1-x}Ge_{x}}$ platform.


\section{Results and discussion}

\subsection{SOT-driven nanomagnets to apply control pulses to the qubit}

The magnetization dynamics of the SOT-driven NM is simulated with the micromagnetic simulation package MuMax3, using the modified Landau-Lifshitz-Gilbert (LLG) equation (see section \ref{Methods: LLG} for full details). The SOT device consists of a Co-Ni multilayered nanomagnet (100 nm × 80 nm × 3.4 nm) placed atop a platinum (Pt) strip (80 nm wide, 4 nm thick) capped with a 2 nm tungsten (W) layer. This assembly is positioned h = 90 nm above the qubit plane (see supp.). In the presence of the external bias field, $B_{0}=200$ mT, the chosen elliptical geometry of the NM, along with the multiple alternative FM layers, allows us to tune the FMR frequency of the stray-fields to the frequency of ESR, by leveraging the in-plane shape anisotropy, and interfacial perpendicular magnetic anisotropy (PMA). With the major axis (easy axis) of the NM aligned parallel to the HM-strip, the eccentricity, and the volume of the former are optimized to achieve the desired operational frequency, and field amplitude.

\begin{figure}[ht]
    \centering
    \begin{subfigure}{0.85\linewidth}
    \centering
        \includegraphics[width=1.0\linewidth]{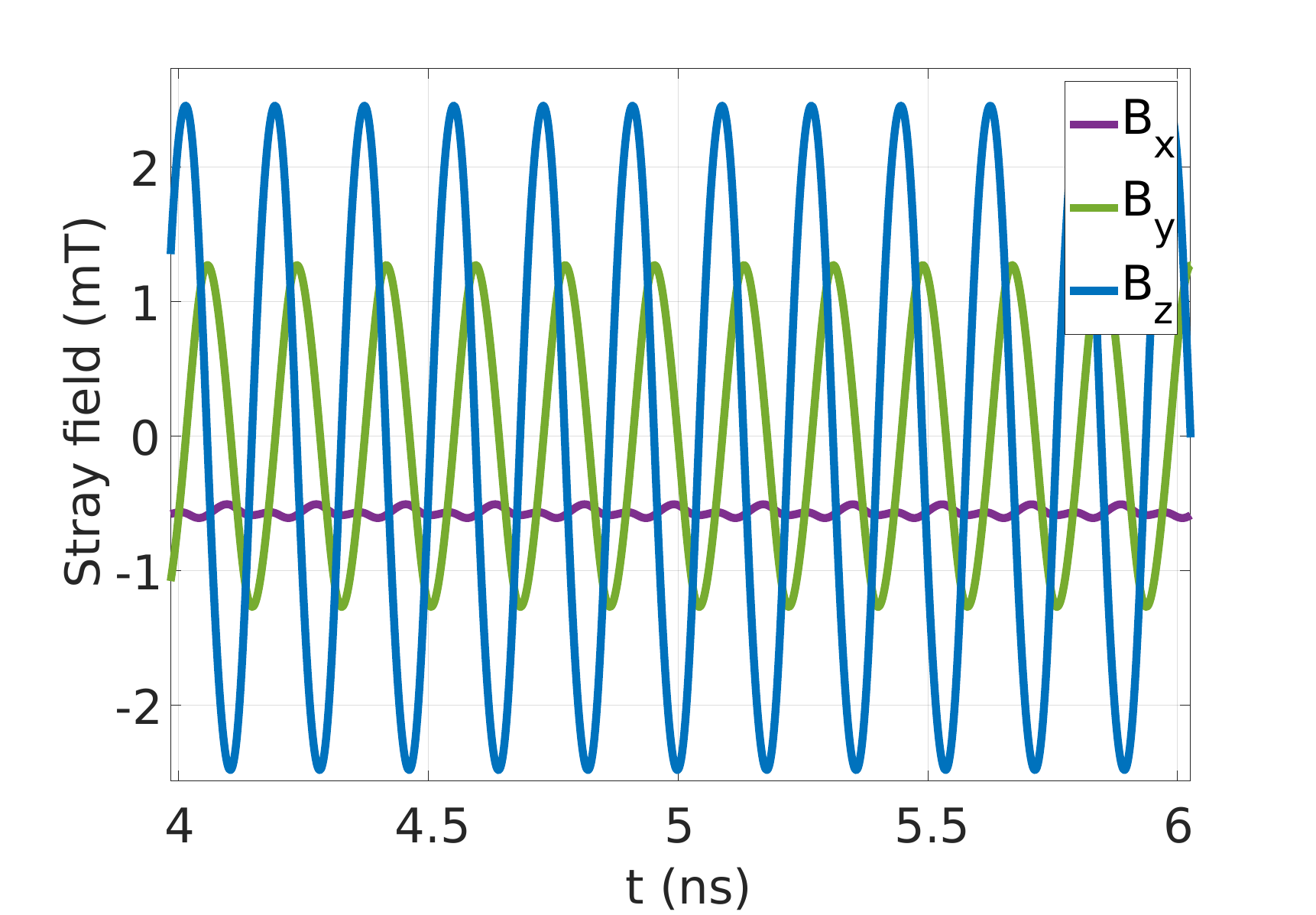}
        \caption{}
        \label{fig:B}
    \end{subfigure}
    \begin{subfigure}{0.85\linewidth}
        \centering
        \includegraphics[width=1.0\linewidth]{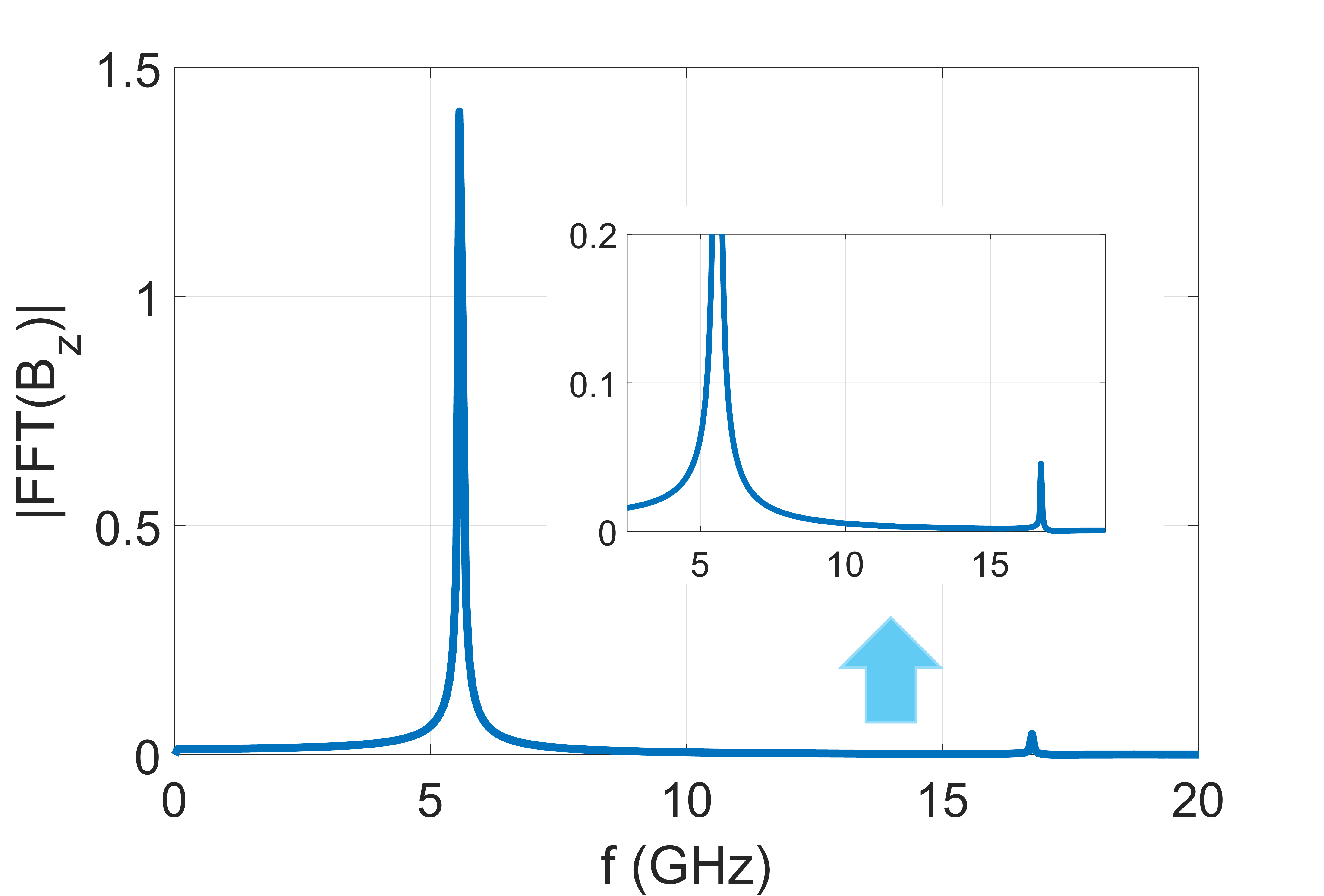}
        \caption{}
        \label{fig:FFT}
    \end{subfigure}

    \caption{\textbf{(a)} The three components of the oscillating magnetic field generated by the precessing magnetization of the NM. \textbf{(b)} The frequency components of the $B_z$ component, with the higher harmonic enlarged in the inset.}
  
    \label{fig:B_F}
\end{figure}

Table \ref{tab:simulation_parameters} lists the simulation parameters used to model the magnetization dynamics of the NM near its FMR frequency. Since the magnetization easy axis is aligned parallel to the $\hat{x}$  direction, applying a time-varying SOT current at the FMR frequency induces oscillating torques that drive the magnetization to precesses around the effective field along the x-axis ($\mathrm{B_0}$). To benchmark the performance of the device, we consider a $200$ $\mu$A charge current, applied along the SOT-strip described earlier (Fig. \ref{fig:SOT}). The three components of the oscillating stray field produced by the NM, along with the frequency spectrum of the $\mathrm{B_z}$ component, is shown in  Fig. \ref{fig:B_F}. 


As the stray field is elliptically polarized, the Rabi frequency is defined as:
\begin{equation}
 f_R = \frac{g \mu_B (B_y+B_z)}{2h},
 \label{eq: Rabi rate}
\end{equation}

where $B_y$ and $B_z$ are amplitudes of the transverse field components that act as the control field for the spin qubit. For a representative configuration, we estimate the stray field components of the NM at the qubit location to be $B_y = 1.27$ mT, and $B_z = 2.464$ mT, which yields a Rabi rate of $52.25$ MHz according to Eq. \ref{eq: Rabi rate}. For comparison, the resonant AC magnetic field generated from the current in the SOT-strip is only 0.4 mT ($=\frac{\mu_0 I}{2 \pi h}$) for the same conditions. Being, linearly polarized, this yields a Rabi frequency of 5.6 MHz in the absence of a NM, demonstrating an order-of-magnitude enhancement over the Oersted field drive in our approach.


The non-zero static stray field ($\approx0.5$ mT) from the NM (Fig. \ref{fig:B}) opposes the external magnetic field ($B_0=200$ mT) at the qubit position, reducing the effective Zeeman field to 199.5 mT, and yielding an ESR frequency of 5.586 GHz. For a given external magnetic field $\mathrm{B_0}$, and effective PMA, the volume and the aspect ratio (major/minor axis ratio) of the NM are chosen such that its FMR equals, or slightly exceeds, the qubit's ESR frequency. When driven by an RF SOT current, the oscillating B-field exhibits a dominant frequency component at 5.586 GHz, along with a few weak higher harmonics (Fig. \ref{fig:FFT}). While a single-mode oscillating B-field is ideal for coherent spin rotation, we demonstrate (see supp.) that the weak higher harmonics have negligible effect on the SR fidelity.

\begin{table}[t]
\centering
\caption{List of parameters used in the simulation.}
\label{tab:simulation_parameters}
\begin{tabular}{l c}
\hline
\textbf{Parameter} & \textbf{Value} \\
\hline
Saturation magnetization ($M_s$) 
& $5.8 \times 10^{5}\ \mathrm{A\,m^{-1}}$\\

Gilbert damping constant ($\alpha$) 
& $0.02$\\

Exchange stiffness ($A_{\mathrm{ex}}$) 
& $1 \times 10^{-11}\ \mathrm{J\,m^{-1}}$\\

Effective PMA constant ($K_{u}$)& $176\, \mathrm{kJ\,m^{-3}}$\\

External bias field  ($B_{0}$)& $200\ \mathrm{mT}$\\

Nanomagnet thickness ($c$)& $3.4\, \mathrm{nm}$\\
 Nanomagnet in plane dimension ($a \times b$)&$(100\,\mathrm{nm} \times 80\,\mathrm{nm})$\\

SOT current frequency ($\nu$)& $5.586\ \mathrm{GHz}$\\

Charge current ($I_c$)& $200$ $\mu$A\\
 Spin Hall angle, $\theta_{SH}$& $0.45$\\
 Pt Stripline thickness ($t_{Pt}$)&$4\, \mathrm{nm}$\\
 W Capping layer thickness ($t_W$)&$2\, \mathrm{nm}$\\\end{tabular}
\end{table}

The colormap depicting $f_R$ at the qubit plane is shown in Fig. \ref{fig:rabi_2D}. The peak of $f_R$ ($\approx 52.25$ MHz) is obtained directly below the NM, where we have positioned the spin qubit in our simulations. The peak of $f_R$ decays in the $\hat{x}$ and $\hat{y}$ directions, with FWHMs of 128 nm and 172 nm. Therefore, well within the capabilities of standard CMOS processes, $f_R$ remains comparable to those achievable by EDSR techniques.

\subsection{Decoherence of Spin Qubits} \label{sec: decoherence}

\begin{figure}[H]
    \centering    
    \begin{subfigure}{0.9\linewidth}
        \centering
        \includegraphics[width=1\linewidth]{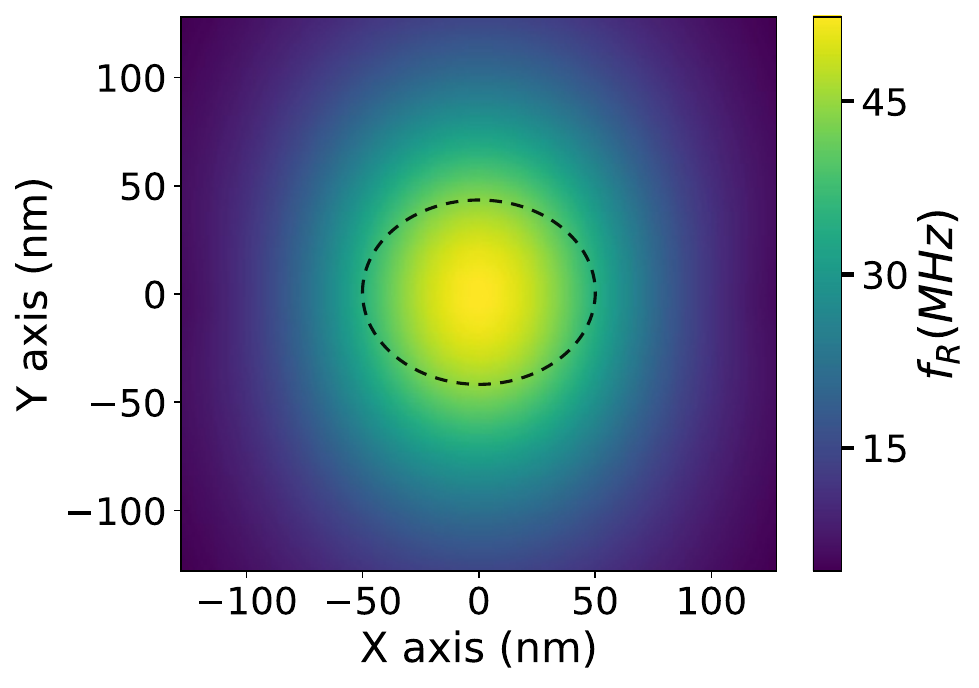}
        \caption{}
        \label{fig:rabi_2D}
    \end{subfigure}
    \vskip\baselineskip
    \vspace{-5 mm} 
    \begin{subfigure}{0.9\linewidth}
        \centering
        \includegraphics[width = \linewidth]{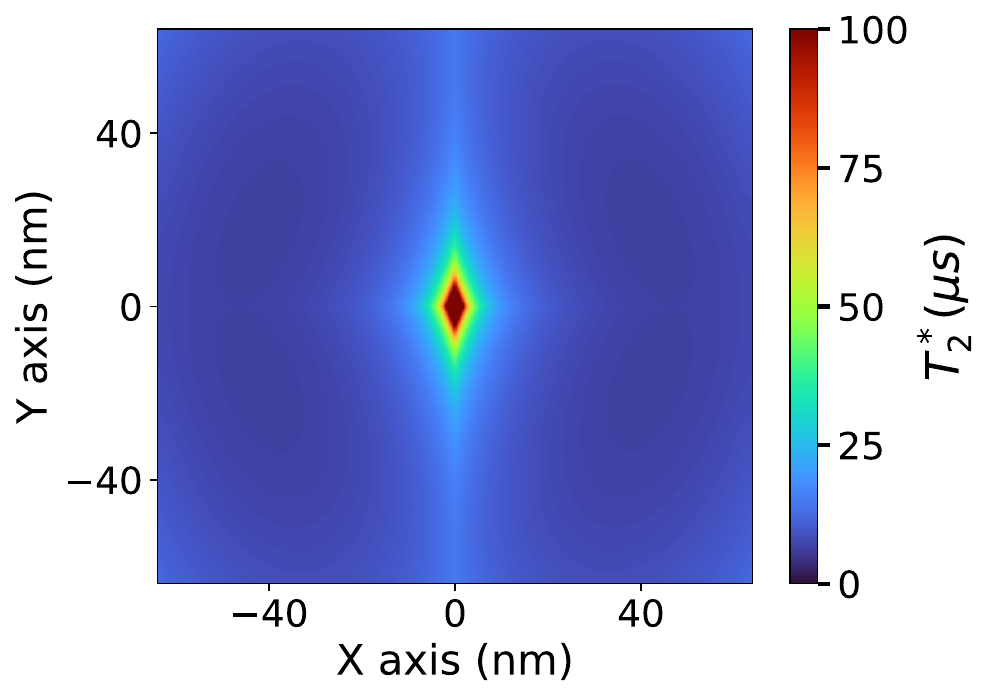}
        \caption{}
        \label{fig:T2star}
    \end{subfigure}
    \vskip\baselineskip
    \vspace{-5 mm} 
    \begin{subfigure}{0.516\linewidth}
        \centering
        \includegraphics[width = \linewidth]{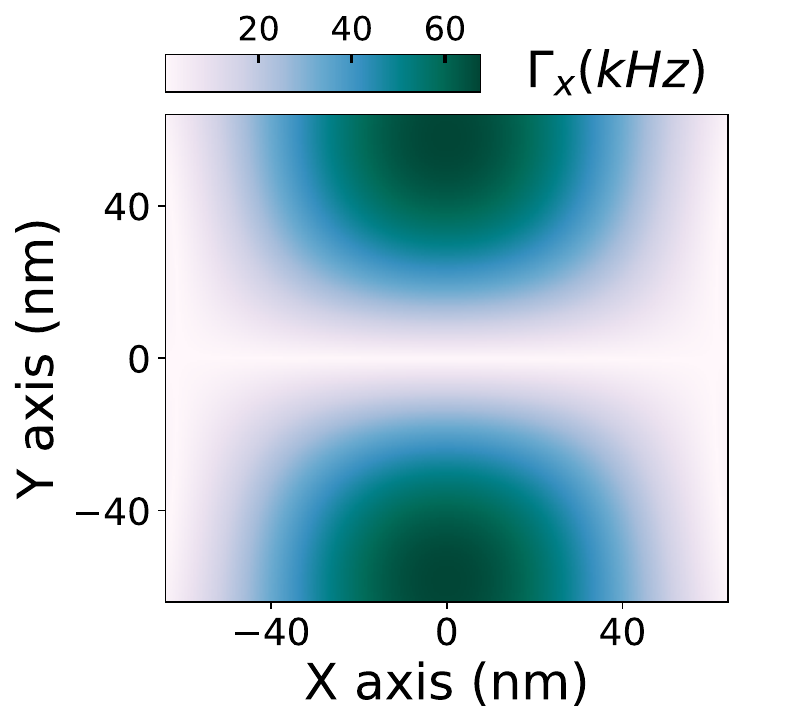}
        \caption{}
        \label{fig: GammaX}
    \end{subfigure}
    \begin{subfigure}{0.426\linewidth}
        \centering
        \includegraphics[width = \linewidth]{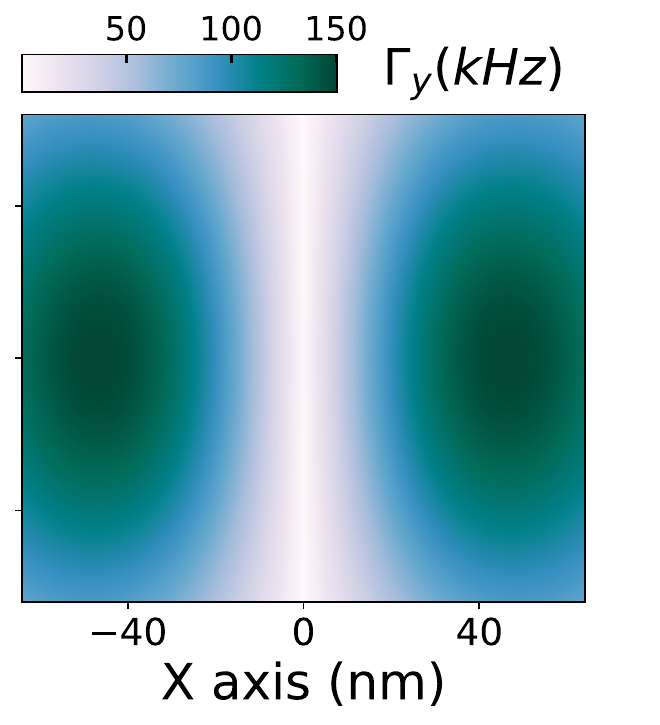}
        \caption{}
        \label{fig: GammaY}
    \end{subfigure}

    \caption{Color-maps of various qubit parameters at the qubit plane. \textbf{(a)} Calculated Rabi frequency at the qubit-site, plotted for a region encompassing the footprint of the entire NM (shown by the dotted ellipse). \textbf{(b)} Plot of $T_2^*=(\Gamma_x+\Gamma_y)^{-1}$ over a smaller region, according to Eq. \ref{eq: T2star}. Plots of the decoherence rate ($\Gamma_x,\,\Gamma_y$) contribution from gradient of Zeeman field along \textbf{(c)} X- and \textbf{(d)} Y- axes respectively.}
\end{figure}

While Fig. \ref{fig:rabi_2D} identifies the region of high transverse field oscillations, which governs the SR speed, the $\mathrm{B_x}$  component of the oscillating field, more appropriately its gradient along the three different axes should be minimized, in order to preserve a sufficiently long coherence time ($T_2^*$) of the targeted qubit. Since a time-varying B-field is directly generated in our proposed scheme, we emphasize here that the B-field gradients are not necessary for the SR implementation (unlike in EDSR), but are inevitable due to the presence of the NM.

The coherence time is calculated as
\begin{equation} \label{eq: T2star}
    T_2^* = \left[ \frac{g \mu_B}{\sqrt{2}\hbar} \sum_{i = {x,y,z}} \frac{dB_x}{di}\Delta_i \right]^{-1},
\end{equation}
where $\Delta_i$ is the root mean square displacement of the spin qubit, under free evolution. Here, we assume a Gaussian decay envelope describing the contribution of the quasi-static component of the noise, with a standard deviation $\sigma = \frac{g \mu_B}{h} \sum_{i={x,y,z}} \frac{dB_x}{di} \Delta_i$, and neglect the high frequency contribution \cite{nakajima2020coherence}. We consider the upper bound of $\Delta_i$ to be 50 pm \cite{kawakami2014electrical} for both in-plane directions, while neglecting the same for the out-of-plane direction (in which the electron wavefunction is tightly bound).

The plots of $T_2^*$ and the decoherence rates, $\Gamma_i= \frac{g\mu_B}{\sqrt{2}\hbar} \frac{dB_x}{di} \Delta_i$, are shown in Figs. \ref{fig:T2star} to \ref{fig: GammaY}. In principle, $T_2^*$ is expected to diverge exactly at the qubit position, making the obtained value of $T_2^* \sim 572 \hspace{3pt} \mu s$ highly sensitive to the simulation-grid size. More robust estimates are obtained at nearby locations: $T_2^* \sim 40.43 \hspace{3pt} \mu s$ and $\sim 95.23 \hspace{3pt} \mu s$ at distances of 5 nm along the x- and y-axes, respectively. Even these conservative values represent a significant improvement over the well-established EDSR scheme \cite{MM_academia1, MM_academia2, maurand_CMOS}. Figures \ref{fig:rabi_2D} and \ref{fig:T2star} together illustrate the tolerance of the proposed layout to qubit placement variability within the qubit plane, thereby assessing the robustness of the SR scheme.

\section{Conclusion}

In conclusion, we propose the use of SOT-driven NMs to locally generate oscillating stray magnetic fields, crucial for rotation of spin qubits hosted by gate-defined QDs in silicon. The technique, which is fully CMOS-compatible, combines high-speed operation with low decoherence rates, and is highly energy-efficient.


While the efficacy of the approach is demonstrated for a single, QD-confined spin, it can be extended for arrays of spin-qubit by careful design of multiple nanomagnets, vis-a-vis the qubit layout. The small footprint of the NM, together with a rapidly decaying dynamic field, is expected to provide site-specific selectivity of qubit rotation, in scaled-up architectures. The proposed scheme is robust against process variations leading to realistic deviations of the QD-positions (5 - 10 nm), and also does not interfere with the gate layout defining the QDs themselves, or the fan-out of the signal routing lines, in the plane below the NMs. Thus, the reported proposal of driving spin rotation can offer a scalable and industrially-relevant mechanism for realizing one- and two-qubit gates for spin quantum computation.


\begin{acknowledgements}
    A.C. and J.A. acknowledge the NSF grant Expand QISE  2231356. K.M., U.M. and S.M. acknowledge the Department of Science and Technology (DST), MoE, for the NQM grant.
\end{acknowledgements}


\section{Methods}\label{Methods}

\subsection{SOT device architecture and micro-magnetic simulation framework } \label{Methods: LLG}
In this device architecture, a multilayer Co-Ni stack has been chosen to fabricate the elliptical NM. This specific geometry provides both perpendicular magnetic anisotropy (PMA), and shape anisotropy, which are required to tune the FMR frequency of the NM. The PMA in a $\mathrm{(Co/Ni)_n}$ multilayer depends on the number of bi-layer repetition n, with PMA generally increasing for $n < 10$, and slowly decreasing subsequently  \cite{PMA_Arora}. The thickness of the individual Co and Ni layers, as well as the seed layer, and the growth conditions can also be varied, to further modify the PMA \cite{PMA_BRAHMA2021}, thus making it highly tunable in this system. However, despite Co and Ni having different values of saturation magnetization ($\mathrm{M_s}$) and PMA ($\mathrm{K_u}$), the multilayer NM stack can be modeled as a single layer, considering the mean values of $\mathrm{M_s}$ ($580$ $kA/m$) and $\mathrm{K_u}$ ($176$ $kJ/m^3$). The multilayer nature of the NM also increases the overall thickness of the NM, thereby enhancing the magnetic moment, and thus yielding large field amplitudes required for a high Rabi rate.

In this setup, the nanomagnet's major axis (x-axis) is aligned parallel to the platinum (Pt) heavy metal (HM)  stripline, and hence parallel to the direction of the charge current. When a charge current flows through the HM layer, the spin-Hall effect generates a pure transverse spin current, which is injected into the NM (along z-axis), with spin-polarization ($\hat{\sigma}$) oriented along y-axis. The intrinsic charge-to-spin conversion efficiency is quantified by the spin Hall angle ($\theta_{SH}$), which encapsulates the strength of the spin-orbit coupling (SOC) of the HM atomic nuclei, the material structure, electrical resistivity, and the HM-FM interface quality (spin transparency) \cite{SHE_Hoffmann2013}. The relationship between spin current ($J_s$) and charge current ($J_c)$ densities is expressed as,
\begin{equation}
    J_s = \theta_{SH} \left( \frac{\hbar}{2e} \right) J_c.
    \label{Eq. js_jc}
\end{equation}
Considering these physical parameters, we select  Pt and W as the HM layer for spin current generation.  To maximize spin current efficiency, the thickness of the primary HM layer (Pt) is chosen to well exceed the spin diffusion length, as the  effective conversion efficiency is given by, \cite{spindiffusion_Kondou_2012}
\begin{equation}
    \eta = \eta_{1}[1 -{sech}(\frac{t}{\lambda_s})].
    \label{eq: spin_diffusion}
\end{equation}
Here, t is the thickness of the Co-Ni layer and $\lambda_s$ is the spin diffusion length. The spin diffusion length primarily depends on the resistivity of the HM and HM/FM interfacial spin transparency. Reported values range from 1.2 to 3.4 nm, primarily determined by the interface quality \cite{spindiffusion_Kondou_2012, spindiffusion_Rojas2014,STFMR_Liu2011}. Since SOT efficiency saturates for HM thicknesses exceeding the spin diffusion length (Eq. \ref{eq: spin_diffusion}), a Pt thickness of 4 nm is selected. To enhance charge-to-spin conversion, the $(Co/Ni)_n$ NM is sandwiched between Pt (4 nm) and W (2 nm), exploiting the opposite signs of their spin Hall angles ($\theta_H > 0$ and $\theta_H < 0$, respectively). The W layer thickness is chosen to lie within the $\beta$- to $\alpha$- phase transition regime, in order to maximize the spin current efficiency \cite{PtCoNiW_Qiu}. The resulting effective spin Hall angle is expected to be $\theta_{SH}=0.45$.

Micromagnetic simulations are performed by applying an RF charge current of amplitude $I_c = 200 \hspace{3pt} \mu A$ through a Pt stripline (of width 80 nm and thickness 4 nm). Due to current shunting, the total current redistributes across the Pt, Co/Ni, and W layers according to their conductivities, as described by Eq. \ref{eq: jstack} \cite{PtCoNiW_Qiu}. 

\begin{align}
    J_{i} =\frac{1}{\rho_{i} } \frac{t_{Pt} + t_{CoNi}+t_{W}}{\frac{t_{Pt}}{\rho_{Pt}}+\frac{t_{CoNi}}{\rho_{CoNi}}+\frac{t_{W}}{\rho_{W}}}J_{stack}
    \label{eq: jstack}
\end{align}

Here, $i \in \{Pt, Co/Ni, W\}$, and $J_{stack}$ is the average current density in the stack. For our simulations, we use the following resistivity values: $\rho_{Pt}=35$ $\mu \Omega\cdot$cm, $\rho_{CoNi}=58$ $\mu \Omega\cdot$cm and $\rho_{W}=35$ $\mu \Omega\cdot$cm, and layer thicknesses: $t_{Pt}=4$ nm, $t_{CoNi}=3.4$ nm, $t_{W}=2$ nm.  The applied average current density in the stack is $J_{stack}\approx264\,\mathrm{GA/m^2}$. Inserting these values in Eq. \ref{eq: jstack}, we obtain $J_{Pt}=J_{W}\approx311\,\mathrm{GA/m^2}$. The NM geometry is optimized to closely match its FMR frequency to the target ESR frequency of the qubit. This is achieved in simulation by iteratively extracting the dominant frequency component of the transverse magnetization, under a Gaussian SOT current pulse, in the presence of an external magnetic field $B_0$. An effective perpendicular magnetic anisotropy of $k_{eff} = 176 \hspace{3pt} kJ/m^3$ ($=2K_s/t_{FM}$), and an aspect ratio (AR) of 1.25 yield an FMR frequency of 5.59 GHz, which is very close to the target ESR frequency.

The RF current flowing through the HM (Pt) layer is converted into a pure spin current of the same frequency. This spin current with polarization $\hat{\sigma}$, exerts an anti-damping torque ($\tau_{DL}$) that drives the magnetization away from its equilibrium position, and primarily governs its resulting precession. Additional contributions to the magnetization dynamics may arise from field-like torques, including those due to the Oersted field of the SOT current stripline, the interfacial Rashba–Edelstein (RE) effect, and the anomalous Hall effect (AHE)-induced torques in the ferromagnetic layer. However, to simplify the analysis, our simulations consider only the damping-like torque from the spin Hall effect (SHE) as the sole mechanism for the effective magnetization precession.  As the magnetization precesses at a certain cone angle, the NM produces an oscillating stray field with a large transverse component, which serves as the control field for the spin qubit.

For the micromagnetic simulation of the nanomagnet, we utilize Mumax3 (v.3.11)\cite{mumax_Vansteenkiste2014}. Under the general spin torque mechanism, the dynamics of the unit magnetization $\mathbf{m}=\mathbf{M}/M_s$ in the nano-scale magnet are governed by the modified  Landau-Lifshitz-Gilbert (LLG) equation. This modified equation includes additional damping-like ($\tau_{DL})$ and field-like ($\tau_{FL}$) torque terms, defined as: \cite{Mumax_Joos2023}.

\begin{equation}
\dot{\mathbf{m}}
= -\gamma\,\mathbf{m}\times\mathbf{H}_{\mathrm{eff}}
+ \alpha\,\mathbf{m}\times\dot{\mathbf{m}}
+ \boldsymbol{\tau}_{\mathrm{DL}}
+ \boldsymbol{\tau}_{\mathrm{FL}} .
\label{eq:llg_sot}
\end{equation}

In Eq. \ref{eq:llg_sot}, the two SOT terms are expressed as,
\begin{equation}
\boldsymbol{\tau}_{\mathrm{DL}}
= \gamma H_{\mathrm{DL}}\,
\mathbf{m}\times(\mathbf{m}\times\mathbf{p}),
\qquad
\boldsymbol{\tau}_{\mathrm{FL}}
= -\gamma H_{\mathrm{FL}}\,
\mathbf{m}\times\mathbf{p}.
\end{equation}
where $H_{DL}$ and $H_{FL}$ are the effective fields, proportional to the applied current. 
In case of the SHE-driven SOT,  damping-like effective field is defined as,
\begin{equation}
H_{DL}
=
\frac{\hbar\,|j_c|}{2 e t_{FM} M_{\mathrm{s}}}\,|\theta_H|,
\label{eq:aj_spin_hall}
\end{equation}

where $t_{FM}$ is the thickness of the ferromagnetic layer (3.5 nm for this specific device); $|\theta_{SH}|$ is the magnitude of the effective spin Hall angle, which represents the combined additive contributions from both the HM (Pt) and capping (W) layers; $M_s$ is the saturation magnetization of the FM (Co/Ni) layer.
In this simulation, we disregard the field-like torque, $\tau_{FL}$, by setting  $\xi=0$ as defined in Eq. \ref{eq:HFL_ratio}.
\begin{equation}
H_{\mathrm{FL}} = \xi \, H_{\mathrm{DL}}
\label{eq:HFL_ratio}
\end{equation}
Using these parameters to numerically solve the modified LLG equation, we accurately model the SHE-driven magnetization dynamics of the NM, and compute the oscillating stray field at the qubit location 90 nm below the SOT device plane. This acts as the control field to coherently drive the qubit. 




\bibliographystyle{plain} 
\bibliographystyle{apsrev4-2}
\bibliography{refs} 

\end{document}


\title{     Supplemental Material\\
Quantum control of spin qubits using SOT-driven nanomagnets}
\author{Aniruddha Chakraborty}
\altaffiliation{These authors contributed equally to this work.}
\affiliation{Department of Mechanical and Nuclear Engineering, College of Engineering, Virginia Commonwealth University}

\author{Kanishk Modi}
\altaffiliation{These authors contributed equally to this work.}
\affiliation{Department of Physics, Indian Institute of Technology Bombay, Powai, Mumbai - 400076, India}

\author{Dhritiman Bhattacharya}
\affiliation{Department of Electrical \& Computer Engineering, Henry M. Rowan College of Engineering, Rowan University}

\author{Uditendu Mukhopadhyay}
\affiliation{Department of Physics, Indian Institute of Technology Bombay, Powai, Mumbai - 400076, India}
\affiliation{Centre of Excellence in Quantum Information, Computing, Science \& Technology, Indian Institute of Technology Bombay, Powai, Mumbai - 400076, India}

\author{Suddhasatta Mahapatra}
\email[Suddhasatta Mahapatra: ]{suddhasatta.mahapatra@gmail.com}
\affiliation{Department of Physics, Indian Institute of Technology Bombay, Powai, Mumbai - 400076, India}
\affiliation{Centre of Excellence in Quantum Information, Computing, Science \& Technology, Indian Institute of Technology Bombay, Powai, Mumbai - 400076, India}

\author{Jayasimha Atulasimha}
\email[Jayasimha Atulasimha: ]{jatulasimha@vcu.edu}
\affiliation{Department of Mechanical and Nuclear Engineering, College of Engineering, Virginia Commonwealth University}

\date{\today}

\maketitle

\section{Placement of the Nanomagnet} \label{supp: Placement}

One can place the proposed NMs with their current stripline at multiple locations with respect to the qubit. A few ways one can place the NMs are (consider qubit to be at (0,0,0) nm):
\begin{itemize}
    \item Placing it away from the electrostatic gates confining the quantum dot in the same horizontal level. We can call this (0,200,60) nm placement. The 60 nm of vertical height is including the 30 nm spacer, and a layer of oxide before the current stripline thickness.
    \item Placing it on top of all the gates but vertically above the qubit. We consider 30 nm spacer, and 60 nm of oxide embedded with gates. This gets us to (0,0,90) nm.
    \item One can also consider etching outside the MESA, and placing the nanomagnet significantly far horizontally while eliminating the vertical distance entirely. To prevent interface of MESA from affecting the qubit quality, we consider it to be at least 500 nm away. This results in (0,500,0) nm placement.
\end{itemize}


The (0,200,60) nm placement is closest to the conventional ESR technique stripline placement, and we initially explored this technique to confirm the advantage of an additional nanomagnet placed over the stripline. While we could achieve significantly higher magnetic field oscillations compared to ESR, the (0,0,90) nm placement turned out to be better in terms of absolute amplitude of oscillations.


\section{Gate Operations} \label{supp: Fidelity}
\begin{figure}[ht]
    \centering
     \nopagebreak
     \vspace{-55pt} 
    \makebox[\linewidth][c]{
    \includegraphics[width=1.5\linewidth]{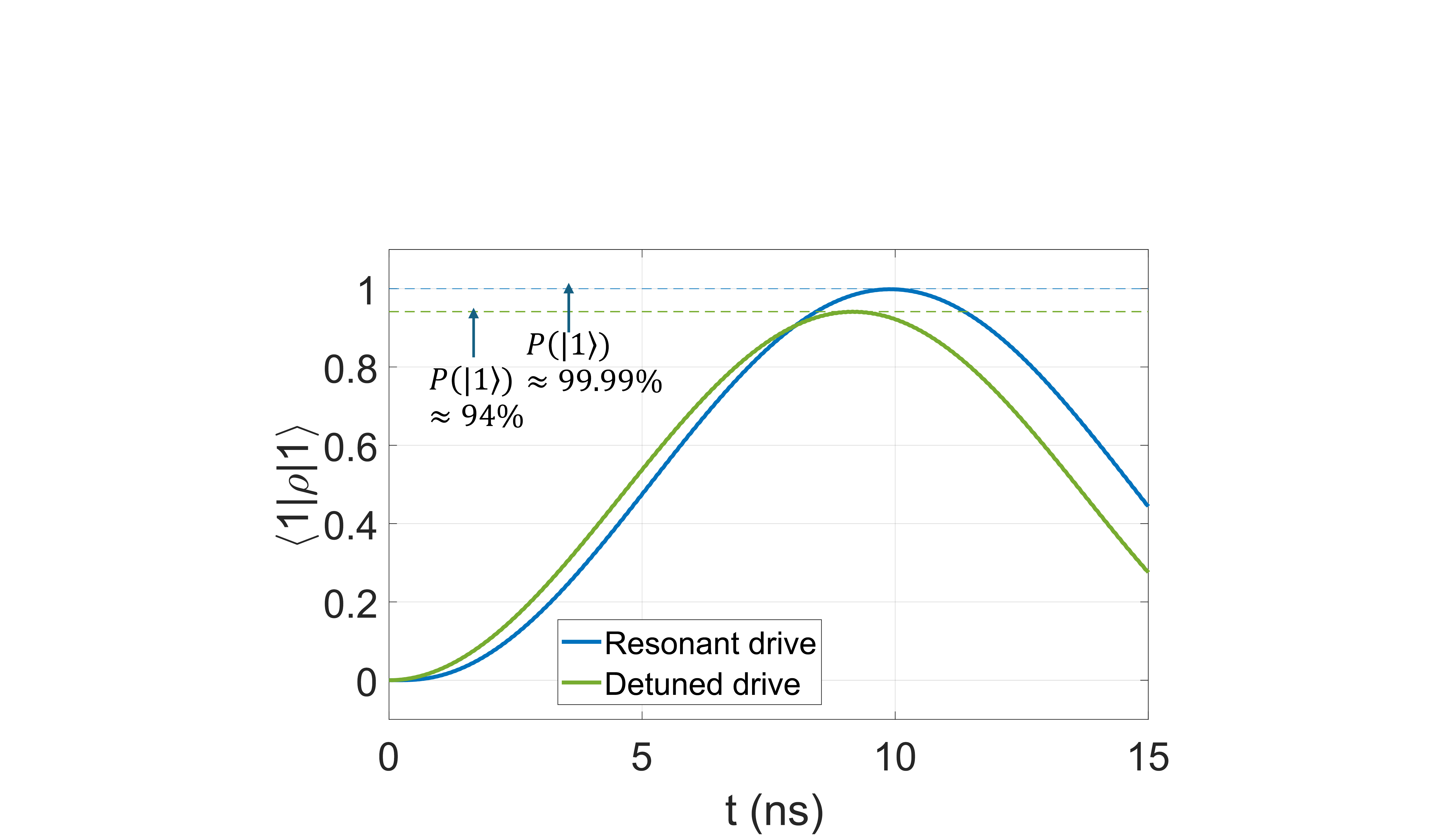}
    }
    \caption{Rabi oscillations driven by the resonant stray fields originating from the nanomagnet, located at (0,0,90) nm with respect to the qubit location, exhibit a gate fidelity of $99.99 \% $, whereas the non-resonant drive (green) shows considerably lower fidelity ($94\%$).}
    \label{fig:fidelity}
\end{figure}

Having established that the SOT-driven NM drive offers a competitive SR rate and long $T_2^*$, compared to EDSR, we now assess the SR fidelity, particularly in the presence of the weak higher harmonics. To implement high-fidelity and fast gate operations, the control field frequency must match the ESR frequency of the spin qubit, whereas the FMR frequency is maintained slightly higher. Once the qubit is initialized, the application of the transverse stray field generated by the NM drives Rabi oscillations.  Accounting for both the external magnetic field ($B_0 = 200$ mT) and the NM stray field ($B_1$), the total Hamiltonian (H) and the corresponding unitary evolution operator (U) in the laboratory frame are defined in eqs. \ref{eq:H} and \ref{eq:U}, respectively). 
\begin{equation}
 \begin{aligned}
  \mathcal{H}
  &= -\frac{\gamma_e\hbar}{2}(B_0\sigma_z+\boldsymbol{B_1(t)}.\hat{\sigma}),
  \end{aligned}
  \label{eq:H}
  \end{equation}
  where $\gamma_e$ is the gyromagnetic ratio of the electron spin, defined as $\gamma_e=\frac{g\mu_B}{\hbar}$;
  \begin{equation}
  \hat{U}(t)
  = \mathcal{T}\exp\!\left\{
   -\frac{i}{\hbar}\int_{0}^{t}
   \mathcal{H}dt'
  \right\},
  \label{eq:U}
  \end{equation}
  where $\mathcal{T}$ is the Dyson time-ordering operator. Using the unitary operator($\hat{U}$) we can estimate the density matrix at each time frame, as defined by,
    \begin{equation}
  \hat{\rho}(t) = \hat{U}(t)\,\hat{\rho}(0)\,\hat{U}^\dagger(t),
  \label{eq:denistymatrix}
  \end{equation}
 where $\hat{\rho}(0)$ is initial state of the density matrix. 
 \\
 Fig. \ref{fig:fidelity} compares Rabi oscillations in terms of spin-state probability under resonant and non-resonant driving conditions. When the NM is driven at the resonant frequency (f = 5.586 GHz), a gate fidelity of 99.99\% is achieved. In contrast, neglecting the static stray field of the NM, and driving the qubit at 5.6 GHz results in a reduced spin-flip probability of 94\%. Unlike Oersted-field-driven schemes, the NM produces a finite static stray field that can limit gate fidelity if not properly accounted for. However, the strong fundamental component and suppressed harmonics of the stray field enable high-fidelity operation, positioning the proposed approach as a viable route towards fault-tolerant qubit control.

\section{Representative Nanomagnet Placement to Increase $T_2^*$} \label{supp: 3nanos}
The Zeeman field gradient causing the spin decoherence can be reduced by optimizing the placement of adjacent nanomagnets. Here, we consider a representative configuration with three identical nanomagnets aligned consecutively along their long axis, with an inter-magnet spacing of $20\,\mathrm{nm}$. Following the decoherence simulation described in the main text, we simulate the magnetic field and calculate $\Gamma_X, \Gamma_Y$ and $T_2^*$.

\begin{figure}[ht]
    \centering    
    \begin{subfigure}{0.9\linewidth}
        \centering
        \includegraphics[width = \linewidth]{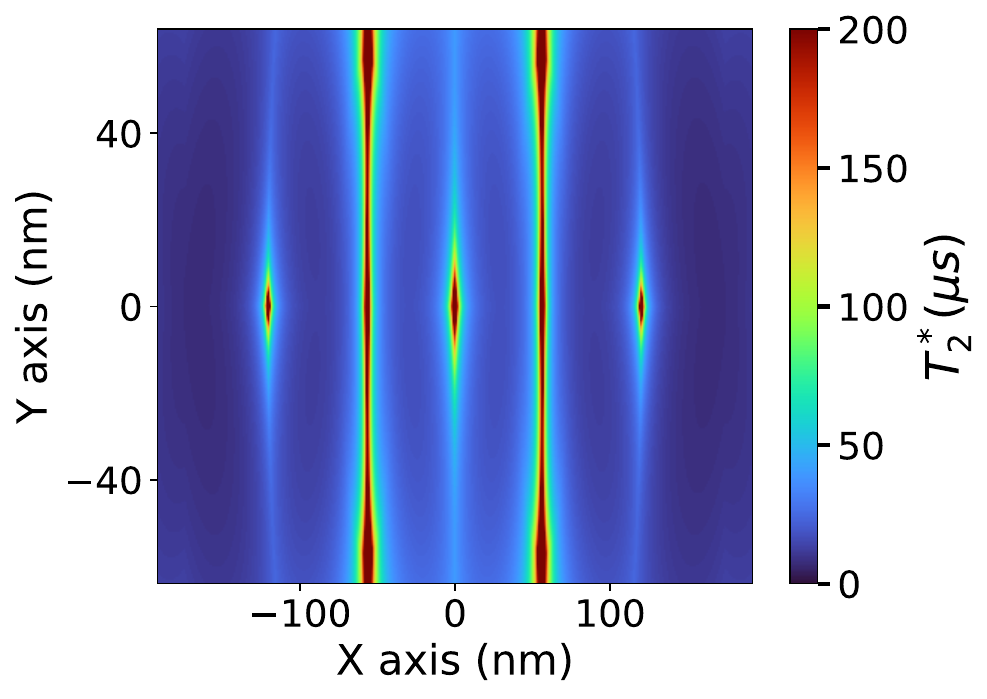}
        \caption{}
        \label{fig:T2star2}
    \end{subfigure}
    
    \vskip\baselineskip
    \vspace{-5 mm} 
    \begin{subfigure}{0.516\linewidth}
        \centering
        \includegraphics[width = \linewidth]{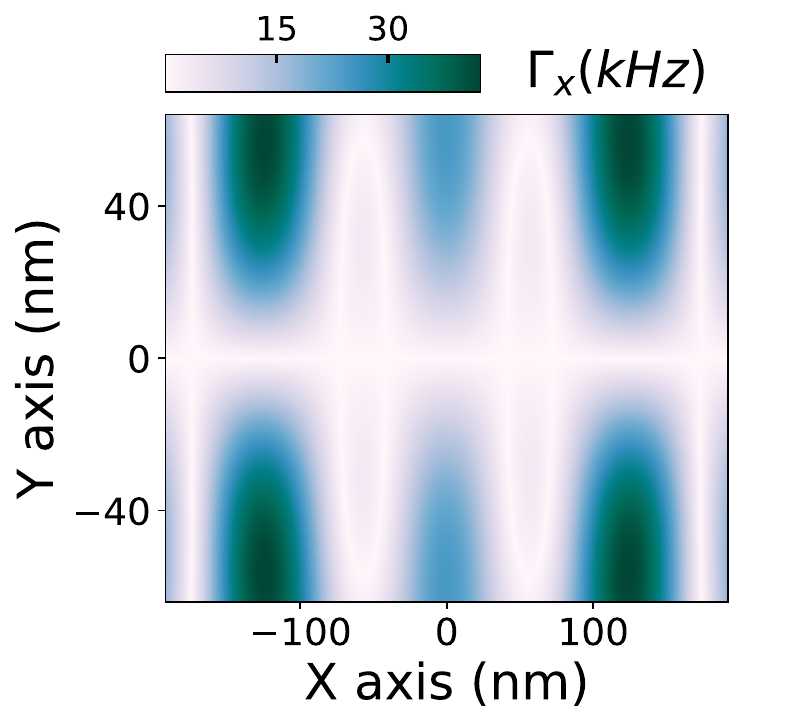}
        \caption{}
        \label{fig: GammaX_3}
    \end{subfigure}%
    \begin{subfigure}{0.426\linewidth}
        \centering
        \includegraphics[width = \linewidth]{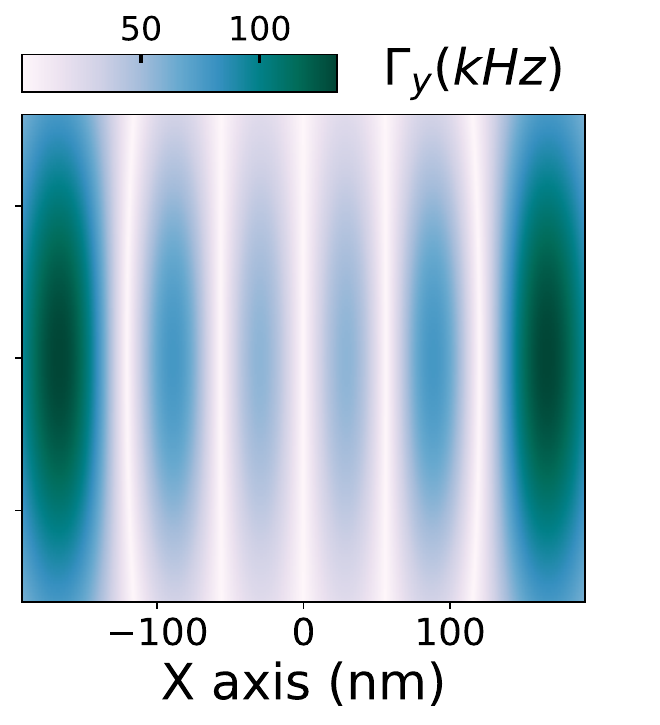}
        \caption{}
        \label{fig: GammaY_3}
    \end{subfigure}

    \caption{Figures show the results for the case of 3 NMs placed next to each other along their major axis with a gap of 20 nm. \textbf{(a)} Plot of $T_2^*$. Plots of the decoherence rate contribution from gradient of Zeeman field along \textbf{(b)} X- and \textbf{(c)} Y- axes.}
    \label{fig: T2star_3nano}
\end{figure}

Simulations with three NMs shows multiple regions of highly coherent qubit operation (see Fig. \ref{fig: T2star_3nano}). As before, the coherence of a qubit placed vertically below the middle NM will not be limited by Zeeman field gradient. This design, even without considerable optimization, achieves $T_2^* \sim 65.04 \hspace{3pt} \mu s$ at (5 nm, 0) position in qubit plane and $T_2^* \sim 252.51 \hspace{3pt} \mu s$ at (0, 5 nm) position. This result emphasizes the coherence advantage of SOT-driven nanomagnets over EDSR for spin rotation, and indicates that scaling the architecture can enhance qubit coherence, contrary to conventional spin-rotation approaches.
\FloatBarrier